\documentclass[page-classic]{epl2}

\title{Holy Memristor}
\shorttitle{Holy Memristor} 

\author{J. Kim\inst{1} \and V.~J. Dowling\inst{1} \and T. Datta\inst{1} \and Y.~V. Pershin\inst{1}}
\shortauthor{J. Kim \etal}

\institute{
  \inst{1} Department of Physics and Astronomy, University of South Carolina, Columbia, South Carolina 29208, USA
}

\abstract{
It was recently shown that when a drop of Glenlivet whisky evaporates, it leaves behind a uniform deposit [PRL {\bf 116}, 124501 (2016)].
We utilize this fascinating finding in the fabrication of electrochemical metallization memory (ECM) cells.
The top (Ag) and bottom (Co) electrodes in our structure are separated by a layer of Glenlivet whisky deposit (an insulator). Measurements show the device response is typical of ECM cells that involve threshold-type switching, pinched hysteresis loops, and a large difference between the high- and low-resistance states. The surface coating process used in our experiments simplifies the device fabrication and results in a biodegradable insulating layer, which may facilitate the recovery of recyclable materials at the end of the device's use.}

\begin{document}

\maketitle

\section{Introduction}
Like the three familiar circuit elements, namely resistors, inductors, and capacitors, memristors~\cite{chua76a} (resistors with memory) are passive two terminal devices. Thanks to their hysteretic response, this newest member of this quadrumvirate of circuitry is able to electronically store  digital or analog information. Consequently, memristors (in the wide sense~\cite{pershin18a}) could potentially be a critical component in the much-anticipated post Von-Neumann revolution in digital computation and electronics.

Starting with the work of Hickmott in the 1960s \cite{hickmott62a}, the memristive effect has been studied in a wide range of solid state devices under a rich variety of material combinations~\cite{pershin11a}. The current research is primarily focused on valance change memory (VCM)~\cite{Sawa08a} cells and electrochemical metallization memory (ECM)~\cite{valov2011electrochemical,menzel2013switching} cells. In the latter, the resistance switching occurs due to the transfer of the active electrode atoms across the  insulator layer under the action of an electric field~\cite{valov2011electrochemical}. Quite often, ECM cells contain RF sputtered silicon oxide as the insulator of choice~\cite{Schindler08a}.

Undeniably, it is of value to explore other approaches to further improve memristor fabrication.
In this article we combine the memristor, the newest passive electrical device, with Scotch whisky~\footnote{Perhaps whisky’s notoriety is exemplified by Pope Francis’ recent comment ‘Questa e la vera acqua santa’ (this is the real holy water). (Rome, 9 October, 2021)}. 
Our work was triggered by reports that well characterized thin films of whisky and related liquids~\cite{Kim16a,saenz2017dynamics} can be produced in the laboratory on various substrates during the natural evaporation of droplets.

Generally, the process of evaporation is quite a complex phenomenon that is influenced by a variety of factors (such as the chemical composition of the drop, substrate type and quality, and environmental conditions).  When a droplet of a simple fluid evaporates, an outward radial flow develops~\cite{deegan1997capillary,deegan2000contact}. This flow replaces the fluid evaporating from the contact line and carries the solute built near the contact line resulting in the coffee ring pattern~\cite{deegan1997capillary,popov2005evaporative}. However, the process of evaporating whiskey droplets  -- a complex fluid containing thousands of organic components~\cite{kew2016chemical} -- is much more complex~\cite{Kim16a}.

In Ref.~\cite{Kim16a}, the formation of a uniform deposit was explained by the interplay of Marangoni flows (necessary for continuous mixing of particles in the liquid) and a strong particle-substrate interaction~\footnote{The Marangoni effect describes mass transfer due to the gradient in surface tension~\cite{marangoni1871ueber,scriven1960marangoni}.}. The process involves several stages. The initial stage is characterized by multiple vortices due to solutal-Marangoni effects created by a concentration variation induced by ethanol evaporation~\cite{Kim16a}. At the next stage, the radial flows develop at the air-liquid interface (outward) and along the substrate (inward). As the evaporation proceeds, this flow pattern  reverses. Such radial flows have been associated with surfactant molecules in whisky (such as natural phospolipids)~\cite{Kim16a}. The absorption of some macro-molecules naturally present in whisky on a substrate has been recognized as another important component of the whole picture that promotes the adhesion and retention of the particles on the substrate. We refer the interested reader to Ref.~\cite{Kim16a} for more information.

As a model mixture, Kim et al. used Scotch whisky (Glenlivet, UK)~\cite{Kim16a}. In a later study,
a similar result was observed using an American whisky with an alcohol-by-volume (ABV) greater than 35\%~\cite{Williams19a}. Moreover, it was found that diluted whisky (ABV$\sim$20\%) leaves a weblike structure, while a very dilute whisky (ABV$\sim$10\%) results in coffee-ring patterns~\cite{Williams19a,Carrithers2020}. Recently, a paper explaining the ‘molecular perspective’ behind improved palatability with water dilution~\cite{karlsson2017dilution} has drawn scientific attention to whisky and other complex fluids.

\begin{figure} [tb]
\centering
(a) \includegraphics[width=65mm]{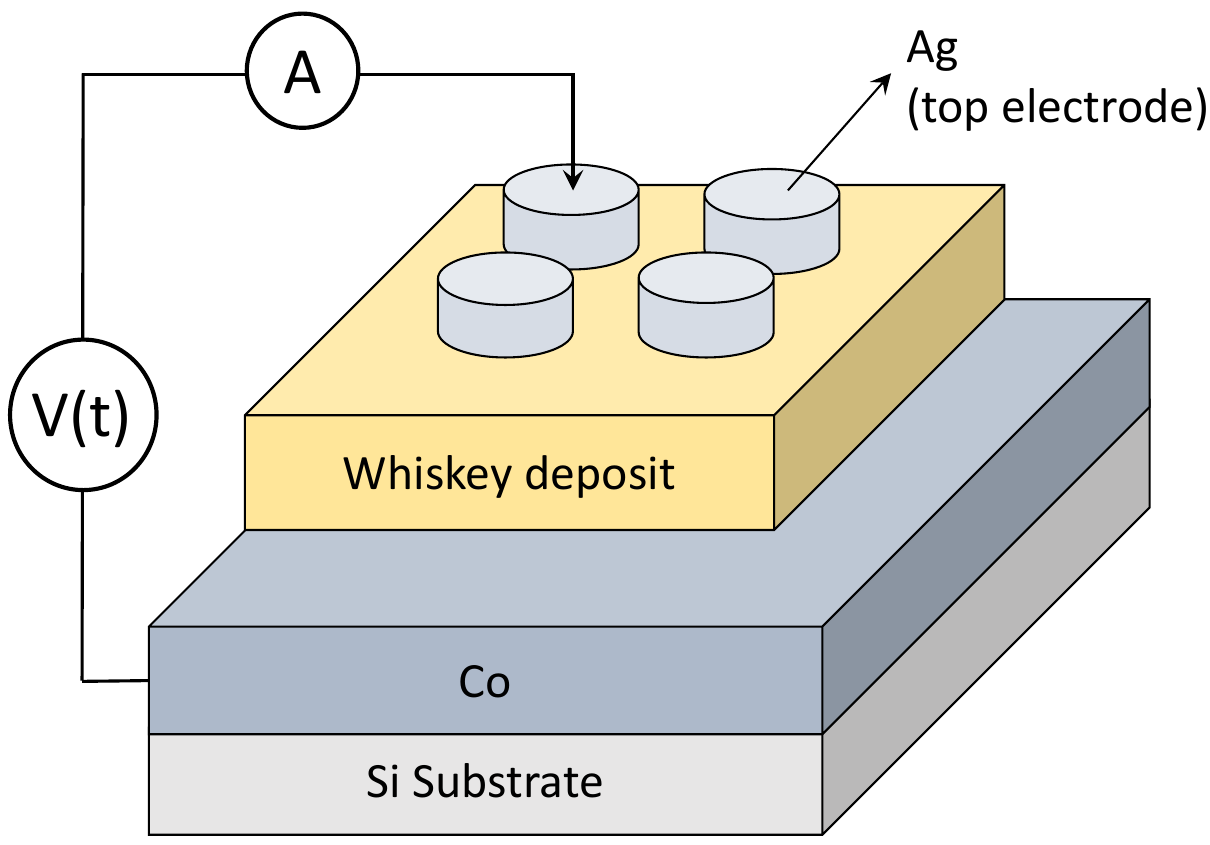} \\
\vspace{0.5cm}
(b) \includegraphics[width=55mm]{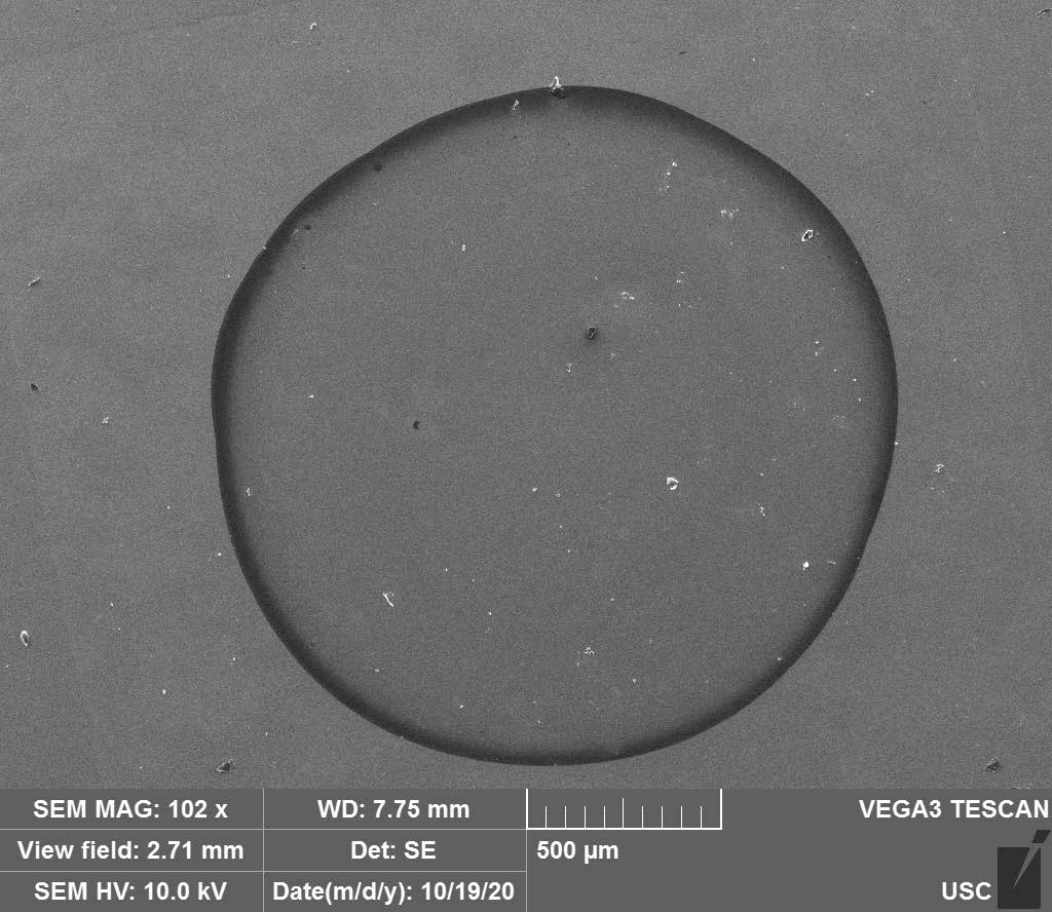}\hspace{0.5cm}
(c) \includegraphics[width=63mm]{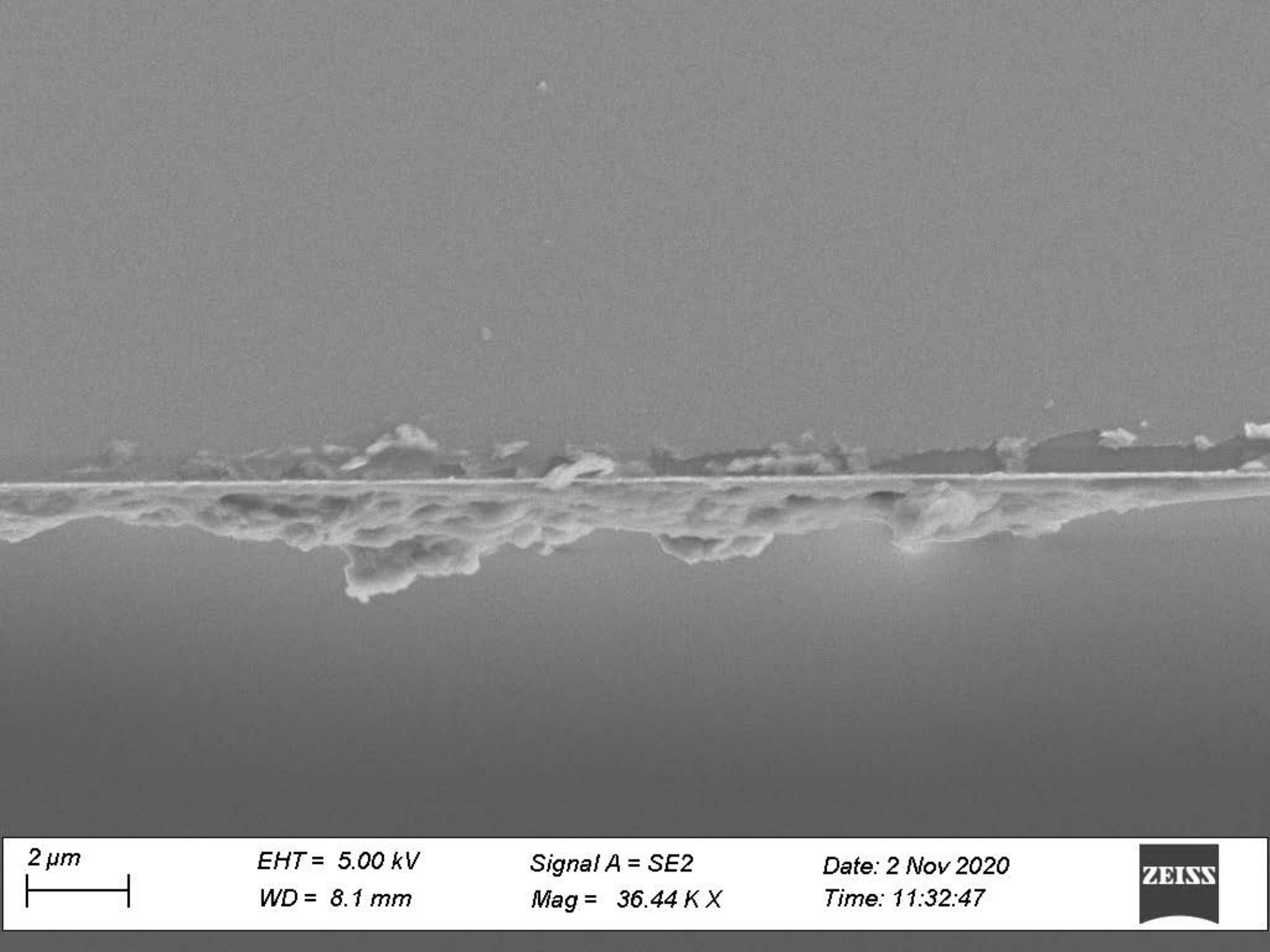}
\caption{(a) Schematic diagram of a Ag/whisky-deposit/Co memory cell and the measurement setup. (b) Top SEM image of a whisky deposit. (c) Side SEM image of whisky deposit. In (c), the straight horizontal line in the middle corresponds to the top edge of the substrate. The thin film of whisky deposit is located above this line. }
\label{fig:1}
\end{figure}

Motivated by the aforementioned results, we introduce and study an innovated memristor structure based on a deposit of Scotch whisky. Such memristors were experimentally fabricated by us, and their properties
were investigated. In particular, we have found that the memory switching characteristics of whisky deposit-based memristors are typical of ECM cells. Our work shows that a whisky deposit can serve as an alternate insulator to the typical insulators used in ECM cells. The replacement of traditional inorganic materials (such as silicon dioxide very often employed in ECM devices) with whisky deposits has a clear benefit in simplifying the fabrication process. Due to the rich physics during film formation of complex fluids, it is possible that these fluids can lead to a more green manufacturing of technology grade films and devices.

The novelty of our contribution is threefold. First, we have simplified the memristor fabrication process by depositing the uniform insulating layer via evaporation of a droplet of a complex liquid. Second, we report the first electronic devices created using (directly) an alcoholic beverage. Third, we report preliminary results indicating that the memristor effect can provide electronic signatures of alcoholic beverages. The fabrication method proposed in this Letter is straightforward, and may be applied to other electronic devices. We also expect that it may have impact in education by bringing the memristor fabrication process to the labs where certain high-vacuum deposition tools are not available.

\section{Experimental details}
\subsection{Fabrication}
Whisky deposit-based memristors were fabricated in a clean room environment following the device structure shown schematically in Fig.~\ref{fig:1}(a). Here, the whisky deposit is sandwiched in between Co and Ag electrodes with the deposit serving as an insulator layer.

\begin{figure}[t]
\centering
	\includegraphics[width=0.45\textwidth]{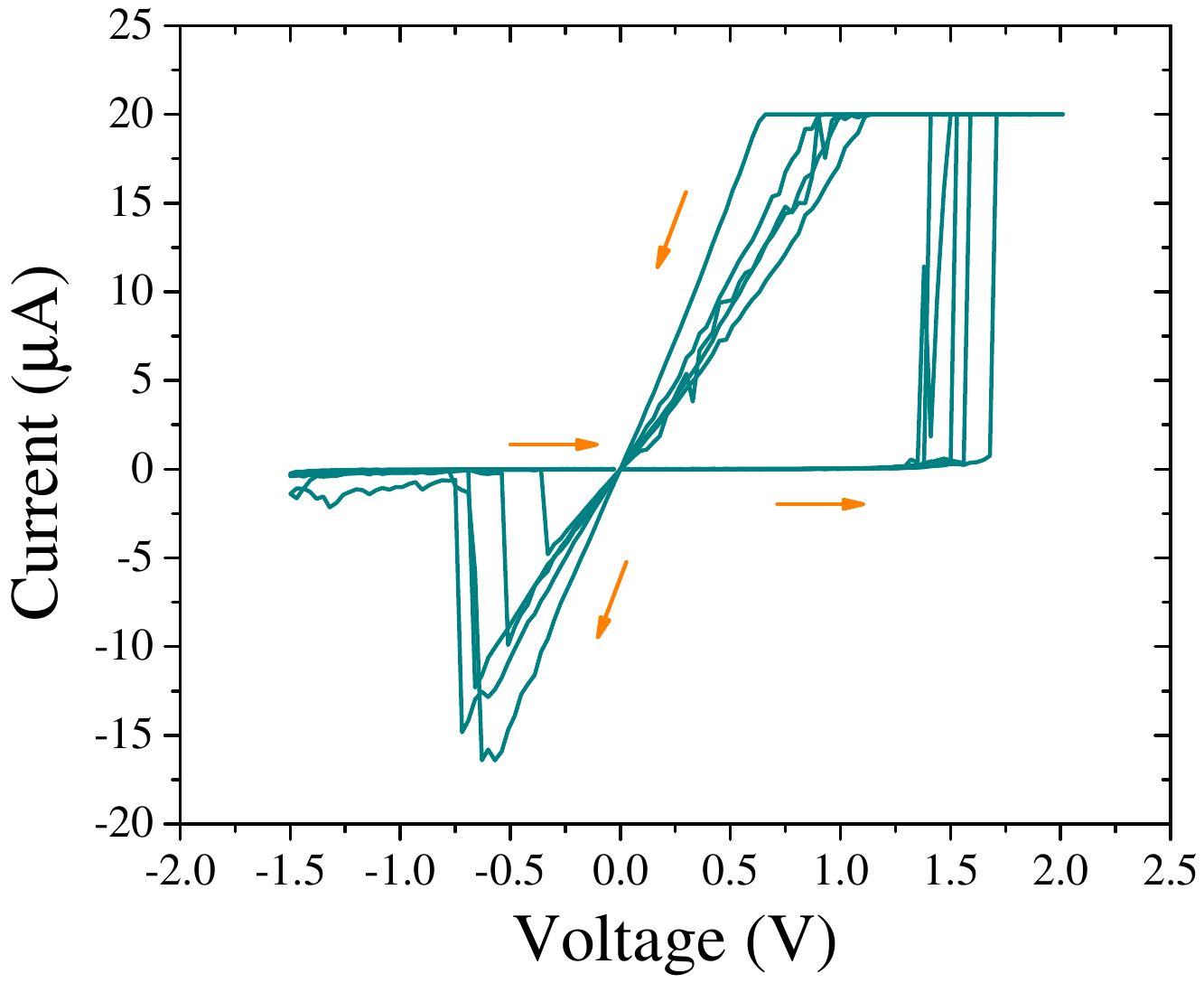}
\caption{Current‐voltage characteristics of Ag/whisky-deposit/Co memristor. Arrows show the direction along the curves.}
\label{fig:2}
\end{figure}

A layer of cobalt (Co) with a thickness of 30~nm (the bottom electrode) was grown by an electron beam evaporator on the surface of an oxidized silicon (Si) substrate.
Roughly 1~cm$^2$ areas on the Co surface were covered manually with Scotch whisky (Glenlivet, UK) that dried under ambient conditions. Finally, the top Ag electrodes (30~nm thickness) were generated on the deposit layer using a metal mask with openings of various shapes and sizes. A cell with a circular top electrode (with a radius of $710$~$\mu$m) that demonstrated reliable switching was selected for the measurements presented in this paper. Multiple memristors fabricated by this method demonstrated similar results to the memristor presented in this paper.

Figure \ref{fig:1}(b) is a top SEM image of the whisky deposit from a small drop showing uniform coverage inside of a circular area. To obtain a side-view SEM image of the deposit (Fig.~\ref{fig:1}(c)), the sample was cleaved across the deposit region. Fig.~\ref{fig:1}(c) confirms the high uniformity of the deposited film, and indicates that its thickness is much smaller than one micrometer (likely about 100~nm or even less).


\subsection{Measurements}
The setup used in this work is typical for the field of resistance switching devices. The memory cells were studied using a probe station connected to a Keysight B2911A precision source/measure unit. In a typical measurement, a triangular voltage sweep
in the range of -1.5~V to 2.0~V was applied to the cell, and the resultant current flowing across the cell was measured. All measurements reported in this paper were obtained with 20~$\mu$A  positive and -1~mA negative compliance currents.

\begin{figure*}[tb]
	\centering
	(a)
    \includegraphics[width=0.45\textwidth]{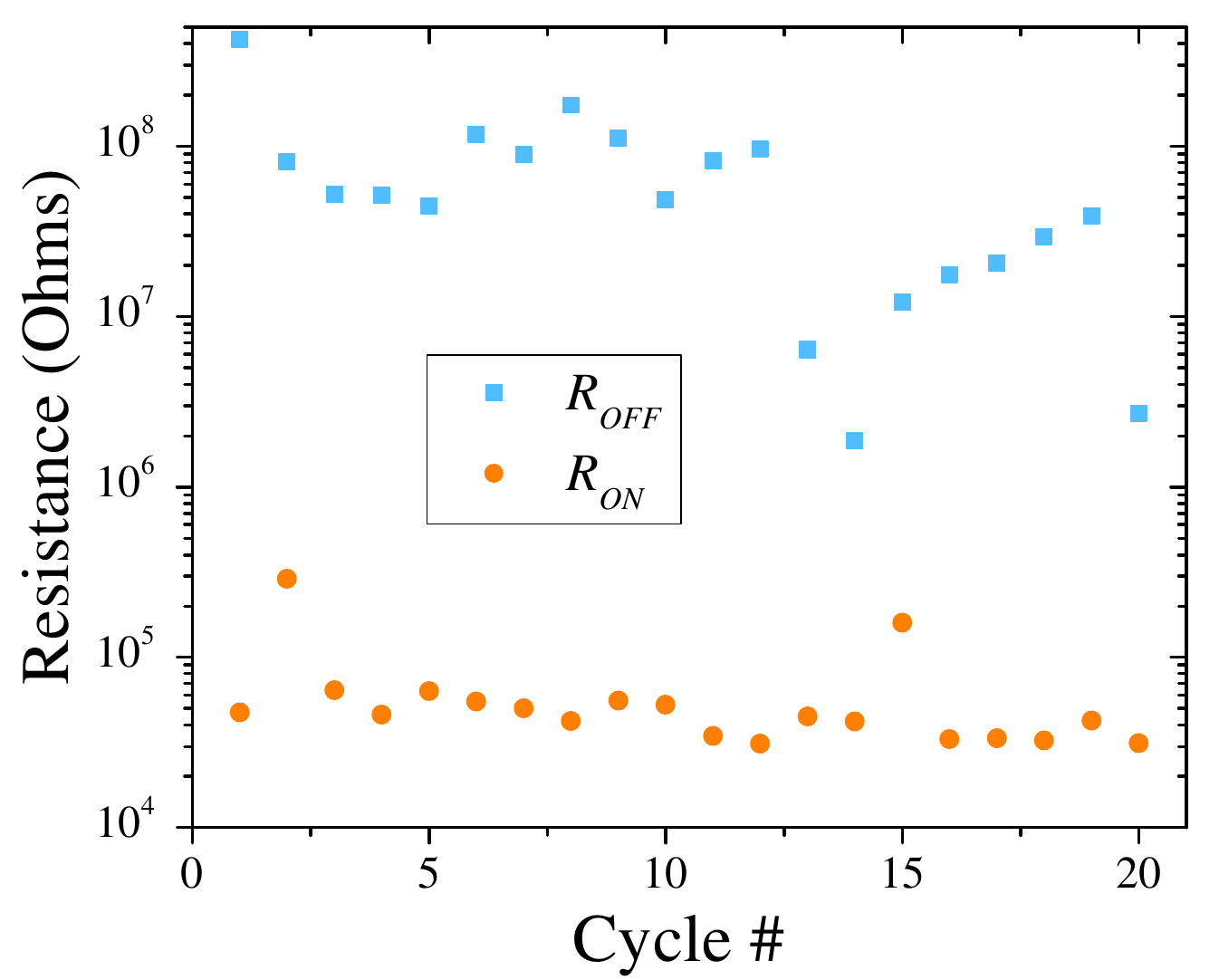}	
	(b)
    \includegraphics[width=0.45\textwidth]{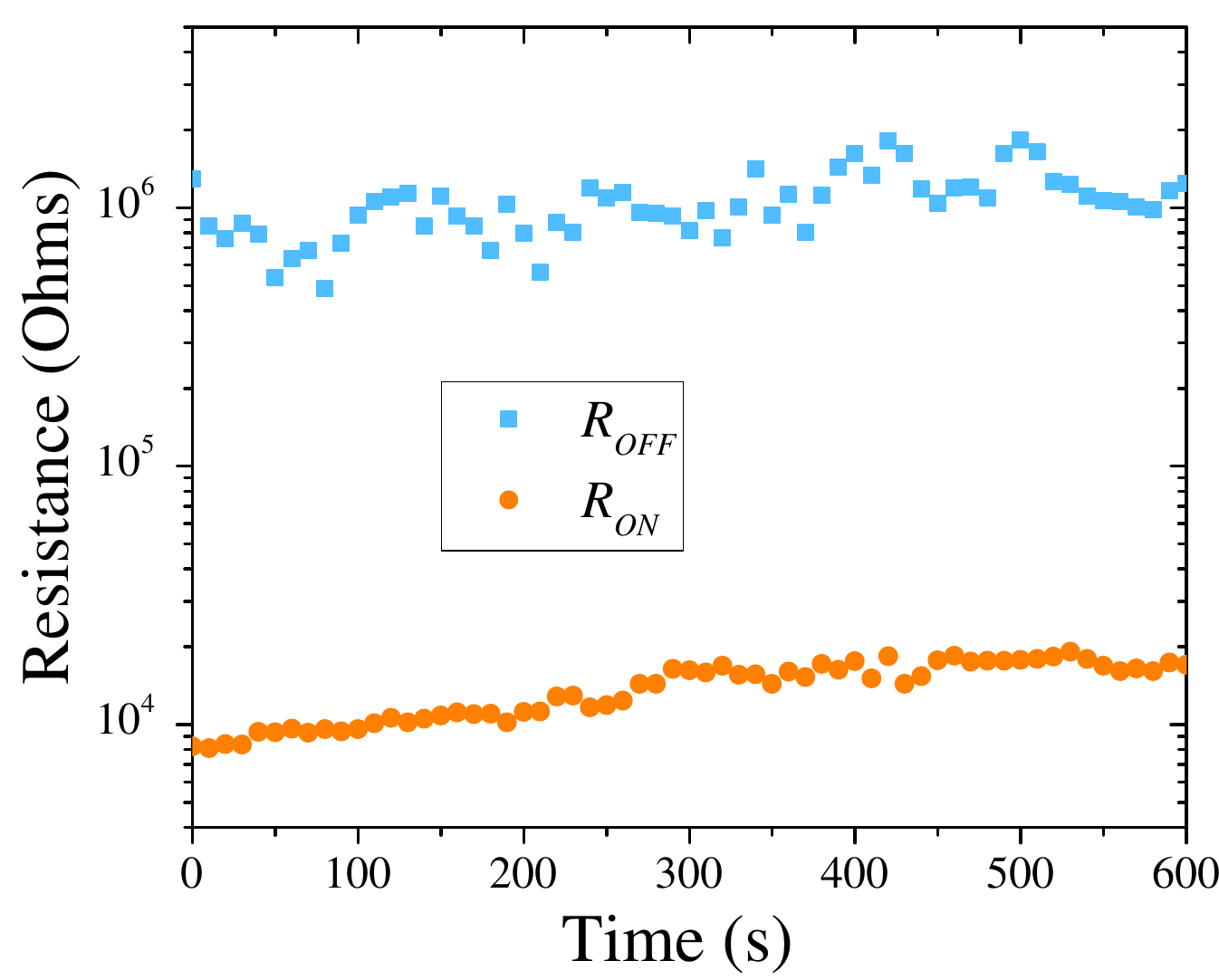}	
	\caption{ Cycle-to-cycle repeatability of resistances in the high- and low-resistance states ($R_{OFF}$ and $R_{ON}$, respectively). (b) The time evolution of $R_{OFF}$ and $R_{ON}$.}
	\label{fig:3}
\end{figure*}

\section{Results}
As in other ECM cells~\cite{valov2011electrochemical,menzel2013switching,onofrio2015atomic}, the anticipated switching mechanism in our devices relies on the formation/dissolution of a metallic filament (composed of Ag atoms) grown across the insulator layer under the action of a high local electric field. When a positive voltage is applied to the active electrode (in our case, Ag), it triggers an oxidation reaction at the electrode/insulator interface. The injected Ag atoms migrate across the insulating layer and are reduced at the insulator/inert electrode (Co) interface. This process leads to the formation of a metalic filament that eventually short-circuits the two electrodes (switching into the ON state). A negative voltage applied to the active electrode induces the migration of Ag atoms towards the active electrode (filament dissolution, switching into the OFF state).

The $I-V$ sweeps presented in Fig.~\ref{fig:2} demonstrate five consecutive cycles wherein the voltage changes linearly in the following sequence: $0$~V$\rightarrow 2$~V$\rightarrow -1.5$~V$\rightarrow 0$~V. In the first step, $0$~V$\rightarrow 2$~V, before the applied voltage reaches the SET threshold voltage ($V_{SET}\sim 1.5$~V), the current flowing across the cell is very small due to the high resistance of the whisky deposit. An abrupt increase in the current at $V\sim 1.5$~V indicates the formation of a conductive filament. The memristor remains in the ON state until the voltage reaches the negative threshold voltage $V_{RESET}\sim -0.5$~V. At this point, the filament dissolves and the electrodes becomes disconnected (the OFF state). The cell stays in the OFF state for the remaining part of the $2$~V$\rightarrow -1.5$~V step and the $-1.5$~V$\rightarrow 0$~V step of the cycle. The cycle then repeats.

In order to evaluate the repeatability of resistance switching, we performed 20 consecutive voltage sweep cycles and extracted  $R_{OFF}$ and $R_{ON}$ from each cycle.
Fig.~\ref{fig:3}(a) shows a very consistent resistance switching 
with a significant cycle-to-cycle variability of low- and high-resistance states. According to Fig.~\ref{fig:3}(a), $R_{OFF}$-s are distributed mainly in between $10^7$~Ohms and $10^8$~Ohms, while $R_{ON}$-s are in between $10^4$~Ohms and $10^5$~Ohms.

Additionally, we investigated the stability of the ON and OFF states over time. For this purpose, the device was initialized into the ON or OFF state, and its resistance was read using low-amplitude voltage pulses at a time step of 10~seconds for 600~seconds. Fig.~\ref{fig:3}(c) proves that the cell has non-volatile information storage capability. We note that Figs.~\ref{fig:3}(b) and \ref{fig:3}(c) were obtained using the same cell with a separation of half a year.


We speculate that along with mass spectrometry and other novel techniques,~\cite{smith2019rapid,Gonzalez17a,rage2020bubbles} the memristive effect and related hysteretic responses can provide tell-tale electronic signatures of alcoholic beverages. In fact, we measured current-voltage curves from samples fabricated using a diluted whisky and observed a correlation between the threshold voltage and level of dilution. The samples used in this study were fabricated as described above with the only difference being that the whisky was diluted with deionized water. The same amount of solution was deposited over the same area to make the insulating layer of memristor.

The positive threshold voltage was measured in three samples that were fabricated with a 1:0, 1:1, and 1:2 whisky to water dilution. The measured values were $V_{th (1:0)}=1.52$~V, $V_{th (1:1)}=1.31$~V, and $V_{th (1:2)}=1.17$~V. These were obtained after averaging the positive threshold voltage measured in 5 consecutive cycles.  This result indicates a tendency for a decrease in the threshold voltage,  $V_{th}$, as dilution increases.  The change in the threshold voltage can be explained by the variation of the deposit thickness (the observed dependence of $V_{th}$ on the whisky content, however, is non-linear). The detailed investigation of this sensing functionality is beyond the scope of the present study.

\section{Conclusion}
Scotch and other whiskys are complex liquids containing thousands of organic components~\cite{kew2016chemical}. In this work, we have introduced a non-traditional application of this important cultural and economical product by showing that deposits from whisky drops can serve as insulating layers in ECM cells. The fabricated cells have demonstrated a reliable switching behavior and non-volatile information storage capability. Compared to inorganic ECM cells, our cells have a larger distribution of resistances in the high- and low-resistance states. This feature, in principle, may be useful in certain applications, such as random number generators~\cite{jiang2017novel,rai2018memristor} and memristive neural networks~\cite{yao2020fully, WEN2018142, pershin09c}.

This study with Scotch whisky films was partially motivated by the universal familiarity of whisky. Our discovery of memristor activity hints at new functionalities
 of custom designed organic films, as well as industrial scale green fabrication of electronic devices.

\acknowledgments
This work was partially supported by an ASPIRE grant from the Office of the Vice President for Research at the University of South Carolina. The authors acknowledge the use of the facilities of the Smart State Center for Experimental Nanoscale Physics at the University of South Carolina.

\bibliographystyle{eplbib}
\bibliography{memcapacitor}

\end{document}